\documentclass[
10pt,
showpacs,preprintnumbers,nofootinbib,
twocolumn,
 amsmath,amssymb,
prl,
groupedaddress,superscriptaddress,
]
{revtex4-1}
\usepackage{graphicx}
\usepackage{bm}
\usepackage[colorlinks=true,urlcolor=blue,citecolor=blue,breaklinks=true]{hyperref}
\usepackage{epsfig}

\newcommand{\cut}[1]{}

\newcommand{\ket}[1]{\ensuremath{| #1 \rangle}}

\begin{document}


\title{Ab initio framework for nuclear scattering and reactions induced by light projectiles}
\author{Konstantinos Kravvaris}
 \affiliation{Lawrence Livermore National Laboratory, P.O. Box 808, L-414, Livermore, California 94551, USA}
\author{Sofia Quaglioni}
 \affiliation{Lawrence Livermore National Laboratory, P.O. Box 808, L-414, Livermore, California 94551, USA}

 \author{Guillaume Hupin}
\affiliation{Universit\'e Paris-Saclay, CNRS/IN2P3, IJCLab, 91405 Orsay, France}
 \author{Petr Navr\'atil}
 \affiliation{TRIUMF, 4004 Wesbrook Mall, Vancouver, British Columbia, V6T 2A3, Canada}

\date{\today}

\begin{abstract}
A quantitative and predictive microscopic  theoretical framework that can describe reactions induced by $\alpha$ particles ($^4$He nuclei) and heavier projectiles is currently lacking. Such a framework would contribute to reducing uncertainty in the modeling of stellar evolution and nucleosynthesis and
provide the basis for achieving a comprehensive understanding of the phenomenon of nuclear clustering (the organization of protons and neutrons into distinct substructures within a nucleus).
We have developed an efficient and general configuration-interaction framework for the description of low-energy reactions and clustering in light nuclei.
The new formalism takes full advantage of powerful second-quantization techniques, 
enabling the description of $\alpha$-$\alpha$ scattering and an exploration of 
clustering in the exotic $^{12}$Be nucleus.
We find 
that the $^4$He$(\alpha,\alpha)^4$He differential cross section computed with non-locally regulated chiral interactions is in good agreement with experimental data.
Our results for $^{12}$Be indicate the presence of strongly mixed 
helium-cluster states consistent with a molecular-like picture surviving far above the $^6$He+$^6$He threshold, and 
reveal the strong influence of neutron decay
in both the $^{12}$Be spectrum
and in the $^6$He($^6$He,$\alpha$)$^8$He cross section. 
We expect that this approach will enable the description of helium burning cross sections and provide insight on how three-nucleon forces influence the emergence of clustering in nuclei. 

\end{abstract}

\maketitle

The low-energy interactions of $\alpha$ particles among themselves and with other nuclei play a prominent role in the helium-burning and later stages of stellar evolution, and are critical to the formation of the chemical elements. Most emblematic is the triple-$\alpha$ process~\cite{Burbidge1957}; the 
fusion of two $^4$He nuclei into the 
transient $^8$Be resonance, immediately followed by the capture of a third $\alpha$ particle. The triple-$\alpha$ process provides one of the main mechanisms for bridging the instability of the mass $A=8$ elements and hence a path for the production of carbon, oxygen and all heavier species.
The description of $\alpha$-induced processes is closely intertwined with the description of nuclear clustering, where
$\alpha$ particles and their dynamical interactions are also the main building blocks, due to their large binding energies. More broadly, elastic scattering and transfer reactions induced by light ($p$-shell) nuclei are often used as indirect means to determine capture reaction rates at stellar temperatures, as well as for probing the cluster structure of exotic nuclear states such as  
molecular resonances in 
beryllium isotopes.

There is a long history of studying $\alpha$-induced reactions and $\alpha$ clustering with microscopic approaches using phenomenological potentials~\cite{Matuse1975,Wada1988,Baye2000,KanadaEnyo2003,IdBetan2012,Funaki2015,Lyu2019,Dreyfuss2020}. Ab initio approaches have only recently begun making first steps, describing $\alpha$-$\alpha$ scattering phase shifts~\cite{Elhatisari2015}, as well as $\alpha$ clustering in carbon and other light isotopic chains~\cite{Epelbaum2012,Elhatisari2017}, starting from nucleon-nucleon (NN) and three-nucleon (3N) forces up to next-to-next-to-leading order of chiral effective field theory (EFT)~\cite{Machleidt2016}. Ref.~\cite{Freer2018} presents a thorough review of microscopic approaches to nuclear clustering.
In this letter we introduce an efficient and general configuration-interaction implementation of the ab initio no-core shell model (NCSM) combined with the resonating group method (or NCSM/RGM approach)~\cite{Quaglioni2008,Navratil2016} that enables the calculation of reactions induced by any light projectile, including $\alpha$ particles and $p$-shell nuclei.  
We present an ab initio calculation of $\alpha$-$\alpha$ scattering with 
chiral NN+3N forces and further demonstrate the versatility of our approach by studying the $^6$He($^6$He,$\alpha$)$^8$He transfer reaction and the cluster structure of the $^{12}$Be system, which has been under thorough experimental~\cite{Korsheninnikov1995,Freer1999,Freer2001,Saito2004,Charity2007,Yang2014,Smith2014,Freer2017} and theoretical~\cite{KanadaEnyo2001,KanadaEnyo2003,Descouvemont2001,Ito2012,Lyu2019} investigation. 

In the NCSM/RGM approach~\cite{Quaglioni2009}, the continuous amplitudes $\gamma_{\nu}^{J^\pi T}(r)$ describing the 
partial wave of relative motion (with spin-parity and isospin $J^\pi, T$) of two nuclei in the reaction channel $\nu$ (specified by the quantum numbers of target and projectile, channel spin $s$ and relative angular momentum $\ell$) 
is found by solving the coupled-channel 
equations  
\begin{align}
\label{eq:inteq}
	\sum_{\nu^\prime}\int dr \, {\mathcal K}^{J^\pi T}_{\nu^\prime\nu}(r^\prime,r)\gamma_{\nu}^{J^\pi T}(r) &= 0\,,
\end{align}
where ${\mathcal K}^{J^\pi T}_{\nu^\prime\nu}(r^\prime,r)$ is a non-local, translationally invariant integration kernel containing all the information about the structure and interactions of the many-body system. We compute the localized (short- and medium-range) components of ${\mathcal K}^{J^{\pi} T}_{\nu^\prime\nu}(r^\prime,r)$ by means of the expansion 
\begin{align}
	{\mathcal K}^{J^{\pi} T}_{\nu^\prime\nu}(r^\prime,r) \!=\!  
	\sum_{nn^\prime} R_{n\ell}(r) R_{n^\prime\ell^\prime}(r^\prime) 
	\sum_{\tilde\beta} {\mathcal M}^{-1}_{\beta\tilde\beta} \,
	{\tilde{\mathcal K}}^{\tilde J^{\tilde\pi}  T}_{\tilde\nu^\prime \tilde{n}^\prime, \tilde\nu \tilde{n}}\,,
	\label{eq:SDchannels}
\end{align}  
where $R_{n\ell}(r)$ is the radial part of a harmonic oscillator (HO) wave functions $\varphi_{n\ell m}(\vec{r})=R_{n\ell}(r)Y_{\ell m}(\hat{r})$, ${\mathcal M}$ is the linear transformation defined in Eq.~(32) of Ref.~\cite{Quaglioni2009} relating relative-coordinate and 
Slater-Determinant (SD) channel states  (with $\beta\equiv\{J^\pi,n,n^\prime,\ell,\ell^\prime\}$), and 
$
{\tilde{\mathcal K}}^{\tilde J^{\tilde\pi}  T}_{\tilde\nu^\prime\tilde n^\prime, \tilde\nu \tilde n} = \left\langle\Xi_{\tilde\nu^\prime \tilde n^\prime}^{\tilde J^{\tilde\pi}  T}\right|H-E	\left|\Xi_{\tilde\nu \tilde n}^{\tilde J^{\tilde\pi}  T}\right\rangle
$
are matrix elements of the microscopic many-body Hamiltonian $H$ minus the total energy $E$ over the reaction channels 
\begin{align}
\nonumber	\left|\Xi_{\nu n}^{J^\pi T}\right\rangle 
	&=\left[ \ket{A_t \, \lambda_{t} I_t^{\pi_t}T_t}_{\rm SD} \ket{A_p \, \lambda_{p} (I_{p}\ell)J_p^{\pi^\prime_p}T_p; n}_{\rm SD} \right]^{J^\pi T}\\
	& = \displaystyle\sum_{i=1}^{N_{\rm SD}} X_{i} a^\dagger_{i_1}a^\dagger_{i_2}\dots a^\dagger_{i_A}|\rangle\,.
\label{eq:DefineXi}
\end{align}
Here, the state $\ket{A_t \, \lambda_{t} I_t^{\pi_t}T_t}_{\rm SD}=\ket{A_t \, \lambda_{t} I_t^{\pi_t}T_t}\ket{\varphi_{00}(\vec{R}^{(t)}_{\rm c.m.})}$ is the SD wave function of the target obtained within the $A_t$-body HO expansions of the NCSM~\cite{Navratil2000,Barrett2013}, 
depending trivially on its c.m.\ coordinate $\vec{R}^{(t)}_{\rm c.m.}$. Similarly,  
\begin{align}
	& \ket{A_p \, \lambda_{p} (I_{p}\ell)J_p^{\pi^\prime_p}T_p; n}_{\rm SD}  \nonumber\\
	& \qquad = \sum_{m_\ell, M_{p}} C_{\ell m_\ell I_p M_p}^{J_p M^\prime_p}B^\dagger_{n\ell m_\ell} \ket{A_p \, \lambda_{p} I_{p}^{\pi_p} M_pT_p}_{\rm SD}
\end{align}
is the eigenstate of the projectile nucleus, obtained once again in the NCSM SD basis, with its c.m.\ wave function subsequently excited in the $\ket{\varphi_{n\ell}(\vec{R}^{(p)}_{\rm c.m.})}$ oscillator state. The formalism and operators $B^\dagger_{n\ell m_\ell} $ for implementing the latter boost of the c.m.\ quanta are described in Refs.~\cite{Kravvaris2017,Kravvaris2019}. Being products of SD wave functions, the basis states of Eq.~\eqref{eq:DefineXi} can be written as linear combinations of $N_{\rm SD}$ $(A_t+A_p)$-body SD wave functions, where $a^\dagger_j$ is a second-quantization operator that creates a particle in a state defined by the set of quantum numbers denoted by $j$, 
and $X_i$ a coefficient determining the weight of the $i$-th SD state in the reaction channel vector, with occupied orbitals $\{i_1i_2\dots i_A\}$. In practice, the total number of Slater determinants for each channel is truncated to make calculations tractable. See Supplemental Material for how this truncation affects observables. 

The present formalism takes full advantage of powerful second-quantization techniques for constructing the reaction channels and evaluating their full matrix elements, unlike our previous approach~\cite{Quaglioni2009,Hupin2013}, where 
only the wave function of the target nucleus and its transition densities were computed starting from Slater determinants.  
Furthermore, compared to 
Ref.~\cite{Quaglioni2009}--where the explicit antisymmetrization of the matrix elements enabled us to distinguish the contributions to the integration kernel coming from the inter-cluster and clusters' Hamiltonians--the present formalism loses track of the identity of particles belonging to the projectile and target nuclei. The full matrix elements of the many-body Hamiltonian are computed at once. 
The two methodologies eventually become equivalent at convergence in the HO expansion. For smaller model spaces minor differences arise in the computed internal energies of the reacting nuclei, which in the 
approach of Ref.~\cite{Quaglioni2009} are included as the eigenenergies at the given NCSM model space size, $N_{\rm max}$. See Supplemental Material for a comparison between the two methods. 
The exact treatment of all other (non-localized) terms of the kernel and the solution of  Eq.~\eqref{eq:inteq} are achieved following the methodology outlined in Ref.~\cite{Quaglioni2009}.

\begin{figure}[t]
\centering
\includegraphics[width=0.47\textwidth]{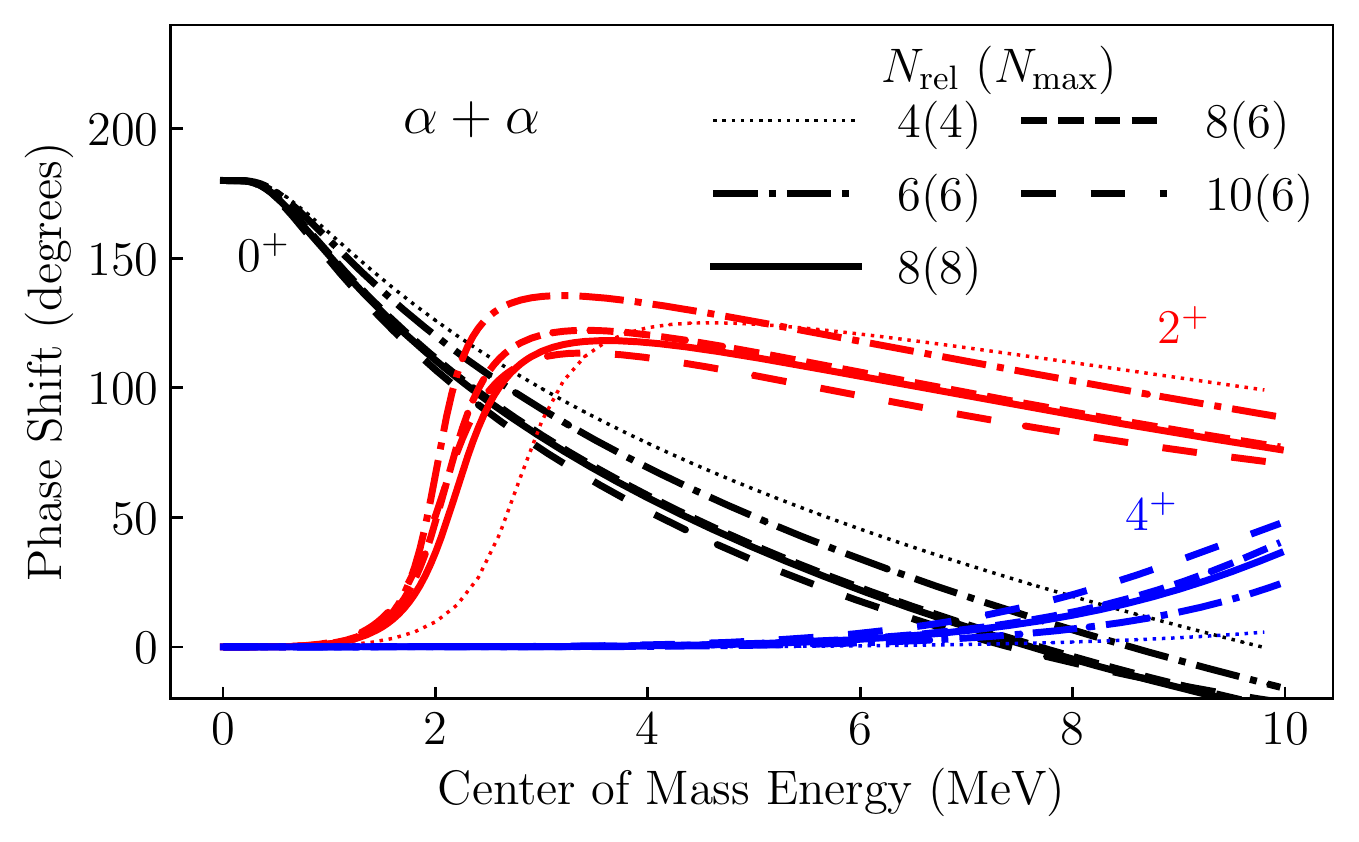}
\caption{
The $\alpha$-$\alpha$ scattering phase shifts obtained with the NN+3N$_{\rm loc}$ interaction reach convergence at $N_{\rm rel}$$\sim$8.
}
\label{fig:comparison}
\end{figure}

As a first demonstration, we  investigate the scattering of two $\alpha$ particles and 
how the 3N force shapes the resulting differential cross section.
The adopted Hamiltonian includes NN and 3N interaction terms:  the N$^3$LO NN interaction introduced in Ref.~\cite{Entem2003} with a regulator cutoff of $\Lambda_{\rm NN}$=500 MeV, denoted simply as NN;
and  
the leading-order 3N force~\cite{VanKolck94} 
in the local form of Ref.~\cite{Navratil2007} with cutoff $\Lambda_{\rm3N}=500$ MeV  of Ref.~\cite{Gazit2019}, denoted as $3\mathrm{N}_\mathrm{loc}$,
as well as  with mixed local and non local regulators 
of Ref.~\cite{Soma2020}, 
denoted as $3\mathrm{N}_\mathrm{lnl}$. In all cases the interactions are softened via a similarity renormalization group (SRG) transformation in three-body space~\cite{Jurgenson2009} with a momentum resolution scale of $\lambda_{\rm SRG}=1.8$ fm$^{-1}$. 
The 3N force matrix elements are included up to a total number of single-particle quanta for the three-body basis of $E_\mathrm{3max}$=16.
\begin{figure}[t]
\centering
\includegraphics[width=0.47\textwidth]{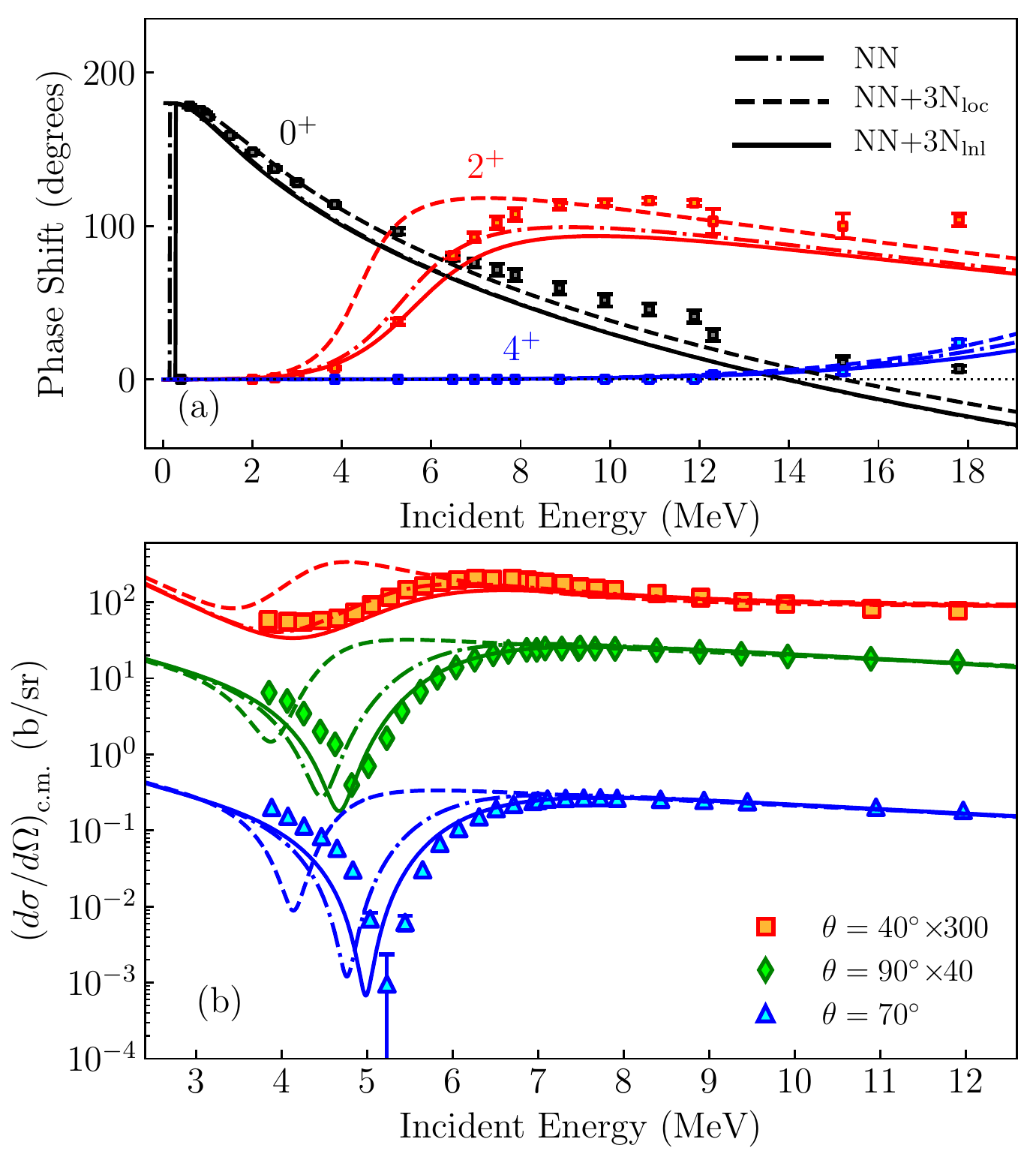}
\caption{The $\alpha$-$\alpha$ scattering (a)  phase shifts and (b) differential cross section at c.m. angle $\theta=40^\circ,90^\circ,70^\circ$, scaled by a factor of 300, 40, and 1 respectively, are sensitive to the choice of chiral interactions: 
NN, NN+3N$_{\rm loc}$, and NN+3N$_{\rm lnl}$ (lines). The NN+3N$_{\rm lnl}$ Hamiltonian yields the overall best agreement with the experimental data (symbols) from (a) Ref.~\cite{Afzal1969} and (b)  Ref.~\cite{Tombrello1969}. } 
\label{fig:dsdo-plot}
\end{figure}

The calculations are carried out with an HO frequency of  
$\hbar\omega$ = 24 MeV and model space size of $N_\mathrm{max}$=8. 
Starting at $N_\mathrm{max}$$\sim$6, where the $\alpha$ wave function 
is close to convergence (Table~\ref{tab:bindingEnergies}), the model-space dependence of the $\alpha$-$\alpha$ phase shifts 
is mainly driven by the convergence of the inter-cluster interaction, that is by the maximum number of quanta in the relative motion $N_{\rm rel}$. 
 A complete $N_\mathrm{max}$=10 calculation would require 3N force matrix elements beyond the current truncation of $E_\mathrm{3max}$=16. 
The convergence error at $N_{\rm max}$=8 can be estimated (while 
still capturing nearly all of the contribution of the 3N force) 
by increasing 
$N_{\rm rel}$ and keeping the overall maximum number of HO shells equal or below 16. 
To this aim, we performed additional calculations with $N_{\rm max}$=6 $\alpha$ particles and $N_{\rm rel}$=8 and $10$. The close agreement between the $N_{\rm rel}(N_{\rm max})$=8(6) and 10(6) phase shifts suggests that convergence is reached around $N_{\rm rel}$=8.
Unless otherwise specified, in the reminder we set $N_\mathrm{rel} = N_\mathrm{max}$. 

The choice of 3N-force regulator plays a major role in determining the 
behavior of the $\alpha$+$\alpha$ system. 
Compared to the results obtained with the NN interaction, where the $^8$Be $0^+$ ground state (g.s.)\ is found as a resonance at 78 keV above threshold, the inclusion of the 3N$_\mathrm{loc}$ interaction produces additional attraction. 
Specifically, the g.s.\ becomes bound by 0.6 MeV and the $2^+$ resonance is shifted to an energy lower than what the phase shifts extracted from experiment~\cite{Afzal1969} suggest (Fig.~\ref{fig:dsdo-plot}a). The lower-energy position of the dip of the differential cross section  is another sensitive indicator of this overbinding. 
Conversely, 
the 3N$_\mathrm{lnl}$ force has a repulsive effect. 
The centroid of the g.s.\ resonance is found at 136 keV above threshold and the dip of the differential cross section moves closer to the experimental data (Fig.~\ref{fig:dsdo-plot}b), yielding the overall most accurate description of the $\alpha$+$\alpha$ system. 
To further demonstrate the general applicability of the approach we now consider the $^{12}$Be nucleus. Its
low-energy continuum is characterized by
multiple 
binary particle decay thresholds. Correspondingly, multiple modes of clusterization need to be included to describe this system. 
We carried out calculations including three of such modes, namely $^{11}$Be+$n$, $\alpha+^8$He, and $^6$He+$^6$He. 
The three-body $^{10}\mathrm{Be}$+$2n$ channel is beyond the scope of this demonstration and is not considered here, though its inclusion can be achieved in the future by combining the present approach with the ab initio frameworks for three-cluster dynamics of Refs.~\cite{Quaglioni2013,Quaglioni2016,Quaglioni2018}. 

As a first application of the approach to reactions with $p$-shell projectiles, we limit ourselves to a soft NN Hamiltonian obtained from the SRG evolution in two-body space of the N$^3$LO NN interaction of Ref.~\cite{Entem2003} with $\lambda_{\rm SRG}=1.8$ fm$^{-1}$, denoted here as NN-only.  The $^{11}$Be and $^{4,6,8}$He wave functions 
are obtained within a $\hbar\Omega=20$ MeV, $N_\mathrm{max}=6$ HO model space.
The parity-inverted ground state of $^{11}$Be results from a subtle interplay of 3N-force and $^{10}$Be+$n$ continuum effects~\cite{Calci2016}, neither of which is included in the present calculations.  
Regardless, we selected the first $1/2^+$ state from the NCSM calculation of $^{11}$Be, despite it not being the theoretically predicted ground state. All helium isotopes are considered to be in their $0^+$ ground states. 
The computed energies for all clusters are summarized in Table~\ref{tab:bindingEnergies}. 

\begin{table}[t]
\begin{center}
\begin {ruledtabular}
\caption{\label{tab:bindingEnergies} NCSM binding energies for the various nuclei used in this work. See text for details about the calculation. }
\begin{tabular}{ c  c c c c c  }
Nucleus & Interaction &	E$_\mathrm{g.s.}$ (MeV)	&$N_\mathrm{max}$ & $\hbar\omega$ (MeV)		 \\
\hline

$^4$He &NN           						&  -25.3     	 &  8 &  24.0      \\
$^4$He &NN+3N$_\mathrm{loc}$           		&  -27.7     	&  4 &  24.0       \\
$^4$He &NN+3N$_\mathrm{loc}$           		&  -28.3     	&  6 &  24.0       \\
$^4$He &NN+3N$_\mathrm{loc}$           		&  -28.4     	&  8 &  24.0       \\
$^4$He &NN+3N$_\mathrm{lnl}$           		&  -28.3     &  8 	&  24.0       \\
$^4$He & NN-only           					&  -28.4     	&  6 &  20.0        \\

$^6$He &NN-only       					&  -27.8 	& 6 &  20.0     	  \\

$^8$He &NN-only 						& -28.6	& 6 & 20.0 	    \\

$^{11}$Be${(\frac{1}{2}^+)}$ &NN-only 		& -65.9	& 6 & 20.0 	    \\
\end{tabular}
\end{ruledtabular}
\end{center}
\end{table}

\begin{figure}[t]
\centering
\includegraphics[width=0.47\textwidth]{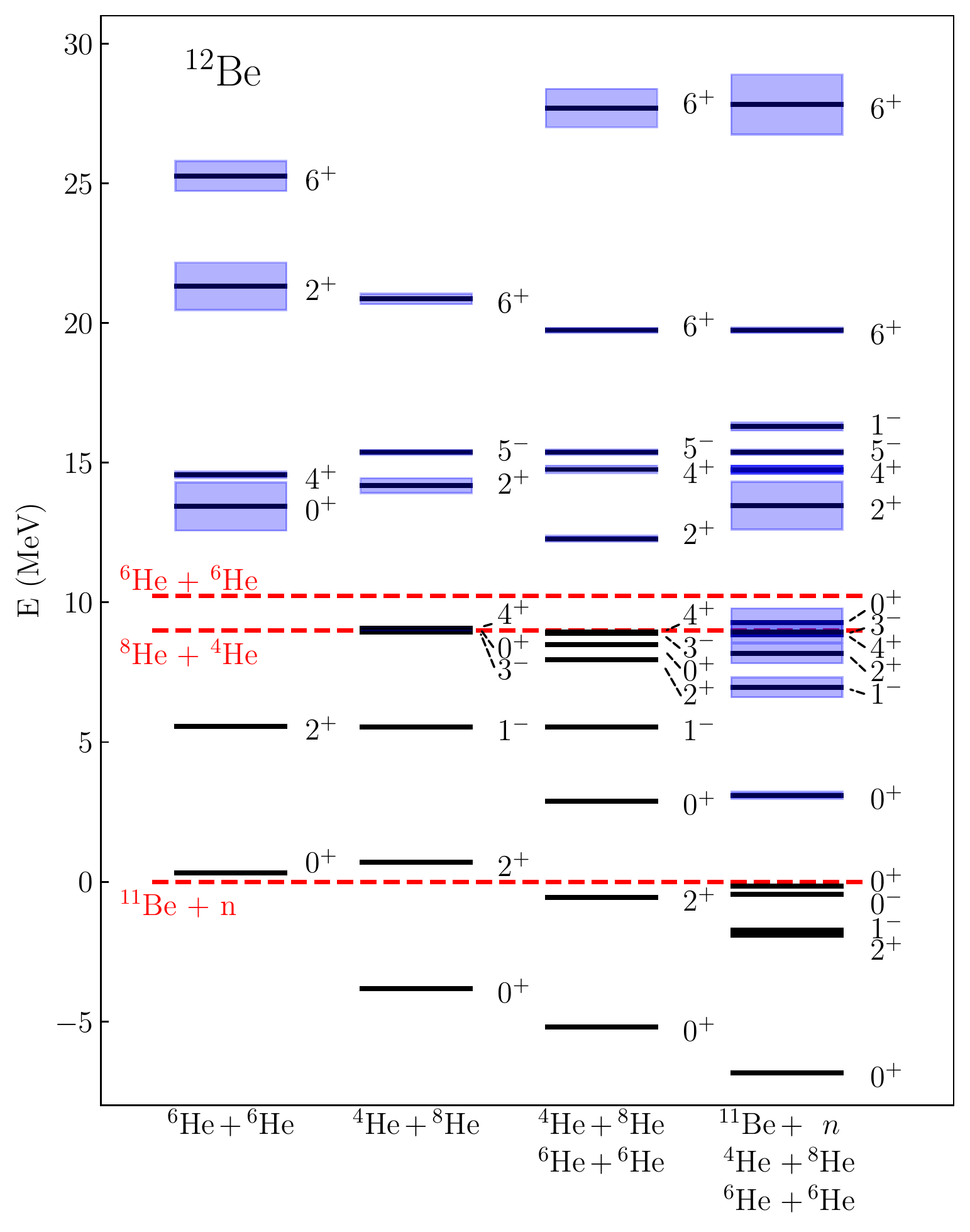}
\caption{Energy levels of the $^{12}$Be nucleus (lines) as a function of the cluster modes included in the calculation. Shaded bands and dashed lines indicate, respectively, resonance widths and the location of the three decay thresholds.} 
\label{fig:Be12Spectra}
\end{figure}

In the limit of no mixing between the $^4$He+$^8$He and $^6$He+$^6$He cluster modes, the NCSM/RGM Hamiltonian would be block-diagonal, and the 
spectrum of energy levels would be the sum of the spectra computed with individual partitions (first and second column of Fig.~\ref{fig:Be12Spectra}). The shift of the energy levels (e.g., $\sim1.3$ MeV for the ground state) in the coupled calculation (third column of Fig.~\ref{fig:Be12Spectra}) suggests otherwise. 
This is consistent with a picture 
of molecular orbits with four valence neutrons covalently shared by two $\alpha$ centers~\cite{KanadaEnyo2003}. A similar result was found in Ref.~\cite{Descouvemont2001}. We note that all negative parity states for the $^6$He+$^6$He system are Pauli blocked, and only natural parity states can exist in both modes. 

The most dramatic change in the spectrum of $^{12}$Be comes from he inclusion of the $^{11}$Be+$n$ channel (fourth column of Fig~\ref{fig:Be12Spectra}). States that were previously bound are now found above the neutron decay threshold. In the absence of 
3N-forces (both chiral and SRG-induced), the calculated g.s.\ energy with respect to the $^{11}$Be+$n$ decay threshold (-6.84 MeV) is overbound compared with experiment (-3.17 MeV). The $0^-$ state appearing just below the decay threshold has not been observed in experiment; it may in fact be the tentative $2^-$ state observed in Ref.~\cite{Smith2014}, though calculations including the $^{10}$Be+$2n$ and $^{11}$Be$(1/2^-)$+$n$ mass partitions would be needed to confirm this assessment. 
The now unbound 0$^+$ state observed around 4 MeV remains essentially constant and does not have an appreciable neutron decay width, 
suggesting that it maintains its helium-cluster characteristics.  
Proximity to a decay threshold, suggested as the underlying mechanism for the emergence of clustering in Ref.~\cite{Okolowicz2012}, does not 
play a role in this case.

\begin{figure}[t]
\centering
\includegraphics[width=0.4875\textwidth]{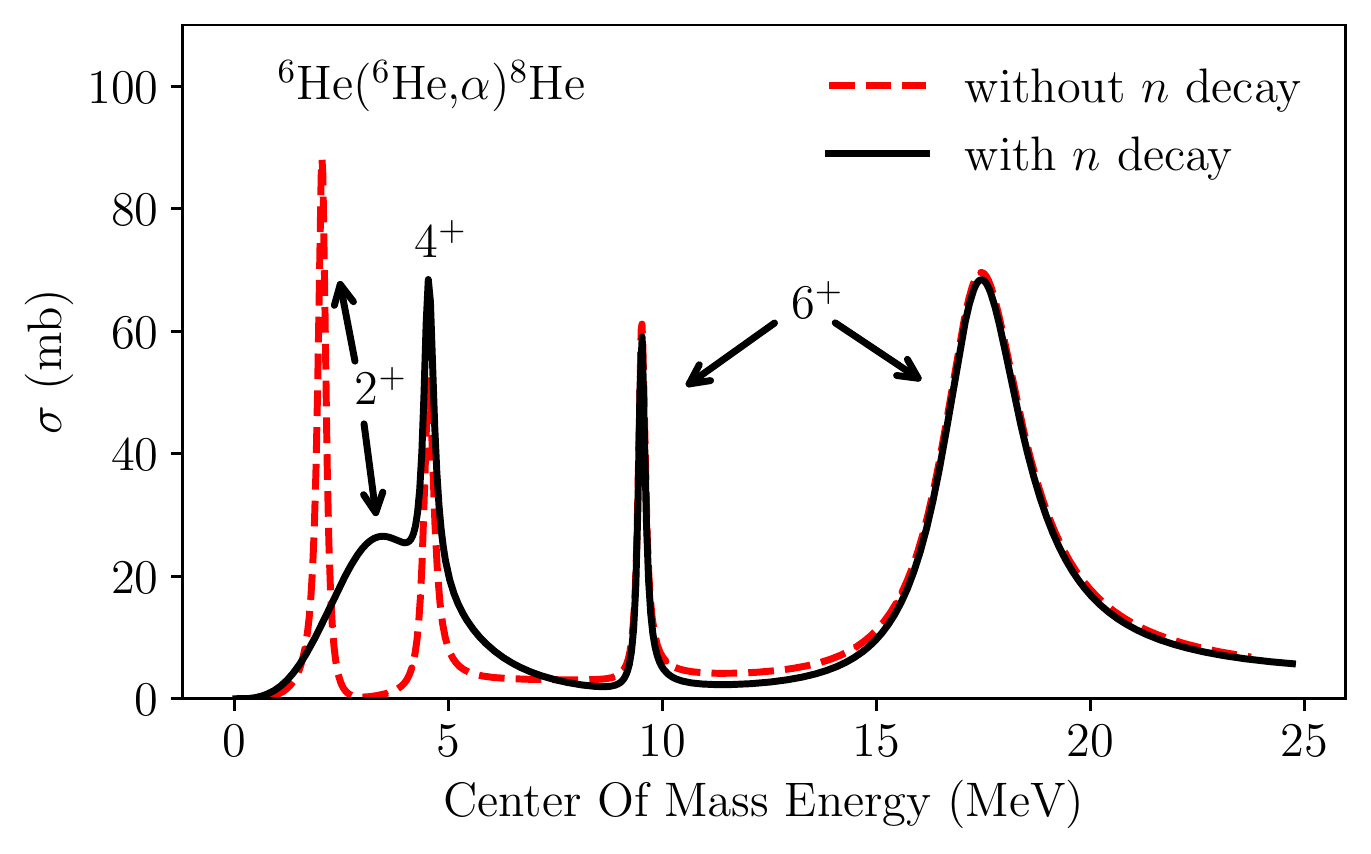}
\caption{$^6$He($^6$He,$\alpha$)$^8$He cross section before (solid line) and after (dashed line) inclusion of the $^{11}$Be+$n$ cluster mode. The $2^+$ resonance acquires a significant neutron-decay width.}
\label{fig:transferReaction}
\end{figure}

In the region above 10 MeV, where states are unbound with respect to the $^6$He+$^6$He threshold, we observe little change with the inclusion of the $^{11}$Be+$n$ channel. That is, nearly all resonances maintain strong helium-cluster characteristics. The sole exception is the 2$^+$ resonance, which acquires a significant decay width. 
This broadening of the 2$^+$ resonance is clearly visible 
in the $^6$He($^6$He,$\alpha$)$^8$He reaction cross section (Fig.~\ref{fig:transferReaction}). This demonstrates that the interaction through the continuum of neutron decay channels allows for further coupling between the two decay modes, leading to a reorganization of the many-body wave function.  Finally, it should be noted that the first 6$^+$ state remains extremely narrow despite lying $\sim$10 MeV from the nearest decay threshold. It can be identified as a helium-clustered state, as it exists in the calculation including both helium mass partitions and remains unchanged with the inclusion of the $^{11}$Be + $n$ channel. 
Such survival of rotational bands well into the continuum 
is an effect of the large angular momentum barrier, which suppresses the decay rate~\cite{Garrido2013,Fossez2016}.

By exploiting c.m.\ quanta boosting techniques in the construction of the microscopic reaction channels used to compute the underlying integration kernel,
we have developed an efficient and general configuration-interaction implementation of the ab initio NCSM/RGM approach.  The new implementation enables 
a major leap forward in the unified ab initio description of nuclear clustering and low-energy reactions induced by any light projectile, including $\alpha$ particles and p-shell nuclei. First applications of the approach reveal the sensitivity of $\alpha$-$\alpha$ scattering observables to the choice of chiral 3N-force regulator, and explain the role of helium clusters and neutron decay in shaping the spectrum of $^{12}$Be and the $^6$He($^6$He,$\alpha$)$^8$He cross section.
%
%
Additionally, the new formalism allows for a controlled lowering of the fidelity of the calculation by truncating the wave functions of the reaction channels at the many-body level. The resulting reduced-fidelity calculations will be a key component of future efforts in quantifying chiral EFT uncertainties for scattering and reaction observables~\cite{KravvarisQuinlan2020}. 
%
The implementation of the present formalism within the generalized microscopic-cluster expansions of the no-core shell model with continuum~\cite{Baroni2013,Baroni2013a}--currently under way--will soon enable the ab initio description of helium burning cross sections and help explaining how 3N forces influence the emergence of clustering in nuclei.



\paragraph{}\begin{acknowledgments}
Computing support for this work came from the LLNL institutional Computing Grand Challenge
program. 
Prepared in part by LLNL under Contract DE-AC52-07NA27344. 
This material is based upon work supported by the U.S.\ Department of Energy, Office of Science, Office of Nuclear Physics, 
under Work Proposal No.\ SCW0498, and by Grant No. SAPIN-2016-00033.  
TRIUMF receives federal funding via a contribution agreement with the National Research Council of Canada. 
\end{acknowledgments}


%

\end{document}